\documentclass[pra,twocolumn,nofootinbib,preprintnumbers,superscriptaddress]{revtex4-1}

\usepackage{natbib}
\usepackage{amssymb,amsbsy,amsmath,amsfonts}
\usepackage{mathrsfs}
\usepackage{graphicx}
\usepackage{epsf,epsfig,float,latexsym,amsthm,fancyhdr,rotating}
\usepackage{graphics,psfrag,longtable}
\usepackage{slashed}
\usepackage{esint}
\usepackage{nicefrac}
\usepackage{braket}
\usepackage{mathtools}

\newcommand{\be}{\begin{equation}}
\newcommand{\ee}{\end{equation}}

\usepackage[colorlinks,citecolor=blue,linktoc=all,linkcolor=cyan]{hyperref} 

\DeclareMathAlphabet{\mathantt}{OML}{antt}{l}{it}
\DeclareMathAlphabet{\mathpzc}{OT1}{pzc}{m}{n}

\def\beq{\begin{equation}}
\def\eeq{\end{equation}}
\def\bea{\begin{eqnarray}}
\def\eea{\end{eqnarray}}
\def\beqa{\begin{equation}\begin{array}{l}}
\def\eeqa{\end{array}\end{equation}}

\def\barr{\left(\begin{array}{c}}
\def\earr{\end{array}\right)}
\def\bmat{\left(\begin{array}{cc}}
\def\emat{\end{array}\right)}

\begin{document}
\title {Recoil Momentum Effects in Quantum Processes Induced by Twisted Photons}

\author{Andrei Afanasev}

\affiliation{Department of Physics,
The George Washington University, Washington, DC 20052, USA}

\author{Carl E. Carlson}

\affiliation{Physics Department, William \& Mary, Williamsburg, Virginia 23187, USA}

\author{Asmita Mukherjee}

\affiliation{Department of Physics, Indian Institute of Technology Bombay, Powai, Mumbai 400076, India}

\begin{abstract}
We consider physical processes caused by the twisted photons for a wide range of energy scales, from optical (eV) to nuclear (MeV) to high-energy gamma-rays (TeV). We demonstrate that in order to satisfy angular momentum conservation, absorption of a twisted photon leads to a transverse recoil of the final particle or a system of particles leading to increased threshold energy required for the reaction to proceed. Modification of the threshold energy is predicted for (a) Photo-absorption on colds trapped ions of $^{40}$Ca, along with emerging new transverse-motion sidebands, (b) photo-disintegration of deuterium and (c) photo-production of electron-positron pairs in astrophysics environment.
\end{abstract}
\date{\today
}
\maketitle


\section{Introduction}


Twisted photons are photons with a shaped wavefront with swirling local momentum or swirling Poynting vectors about a vortex line~\cite{Andrews-book,FrankeArnold_2017}.  
Due to the swirling wave vector, the intrinsic total angular momentum (AM) of the twisted photon along the direction of propagation is $m_\gamma \hbar$, where $m_\gamma$ can be any integer. Processes initiated by twisted photons follow enhanced AM selection rules~\cite{Afanasev:2013kaa,2014PhRvA..90a3425S} different from plane-wave photons. These selection rules have been confirmed by experiments with cold trapped $^{40}$Ca ions~\cite{2016NatCo...712998S, Afanasev_2018}. 

The swirling local momentum of the twisted photon can give significant transverse momentum to the final state, as pointed out by Barnett and Berry~\cite{Barnett_08,Barnett_2013}.   Near a vortex in a monochromatic light beam, the length of the local wave vector, or local momentum, can in fact exceed the wavenumber of any of the plane waves in the superposition representing the beam.   These large  transverse momenta potentially impart what Barnett and Berry call  ``superkicks''  to small particles located near the vortex, as those particles absorb light from the beam. 

It has been explicitly shown in a quantum formalism of twisted photon absorption by single atoms, that the AM that does not go into internal electronic excitations is passed to the target atom's CM motion~\cite{Babiker2002,Afanasev:14} due to AM conservation. Thus the superkick follows as a result of AM conservation. The existence of the superkick, which adds to the kinetic energy of the final state, must lead to a modification of the threshold energies needed for a variety of physical processes.

In the present paper, we consider the kinematics of twisted-photon absorption, on an atom or on another photon, and in particular  how the energy threshold requirements vary with the distance of the target from the photon's vortex line.  We will discuss the significance of the enhanced threshold requirements, and possible effects of upon the reaction cross section.    We will see that in an atomic situation the superkick effect is small but potentially laboratory observable. The effect becomes more pronounced for  processes in nuclear physics and and some cases becomes striking for astrophysically interesting high-energy photon-photon collisions.

Regarding the observability of the superkick, consider an atomic ion struck by visible light twisted photons.   Approaching the vortex line, the density of the photon state decreases.  However, the local momentum relative to the probability density in the same region can get very large.  There is thus a region where densities are very low and the momenta very high.  A sufficiently small probe, for example the ion, fitting in this region may interact rarely but upon interaction will receive a lot of transverse momentum, in some circumstances considerably more than the longitudinal momentum of the Fourier components of the twisted photon.   Hence the name ``superkick.''

While an atom itself is small relative to the wavelength of visible light twisted photons,  the size that matters in the scale of the confinement region in the trap that is holding the ion in place.  That means that the relevant atomic size scale is of order tens of nanometers rather than tenths of nanometers.  Nonetheless the confinement region appears small enough to see an effect, as we shall argue below.    The ion is trapped and the superkick is not sufficient to free it,  but can be enough to push the ion into a higher level in the confining harmonic oscillator potential, with visible consequences.  

An interest in the astrophysical situation lies in the fact that using known physics and unpolarized or simply polarized light, observational estimates of extragalactic background starlight (EBL) give not more that could be obtained from existing visible galaxies.  There had been some expectation that early extragalactic stars existed, and though they no longer exist today, their light would linger in the universe.  The astrophysical data then may be interpreted as either that these early stars never existed or that the universe is more transparent than supposed in the EBL estimates.

The estimates of EBL come from observations of Very High Energy (VHE) photons, or $\gamma$-rays, from distant sources \cite{Aharonian_2006,Albert_2008,Madejski_2016}.  VHE  $\gamma$-ray propagation is diminished by $\gamma\gamma\to e^+e^-$ interactions with EBL,  with $\gamma$-rays having energies above $\approx$ 100 GeV interacting with visible light to produce electron pairs.  Comparing the fluxes of VHE photons as well as lower energy photons from distant sources to the relative fluxes from similar nearby sources allows  an observational estimate of the EBL.   

If there be transparency, the existence of beyond Standard Model axion-like particles, or ALPs, has been offered as an explanation \cite{deAngelis_2007,Jaeckel_2010,Anantua:2010zz}.  The transparency mechanism is that some photons oscillate to ALPs which propagate unhindered and oscillate back into photons.  

Twisted light can be an alternative explanation.  While twisted light is commonly produced on earth, for the present considerations more interesting is that there are mechanisms that produce twisted light in extreme astrophysical situations.  Examples are nonlinear inverse Thompson scattering \cite{Taira_2018},
light bremsstrahlung from energetic electrons as they spiral in strong magnetic fields \cite{Katoh_2017,2019NatSR...9...51M,Maruyama_2019,Maruyama:2019bin}, or by radiation from the warped space near a rotating black hole \cite{Tamburini_2011}.  The reduced cross section engendered by the sometimes higher energy thresholds of twisted photon reactions will lead to increased transparency.

Other novel kinematic effects in collisions of twisted particles have been recently discussed in the literature \cite{Ivanov:2019vxe,Ivanov_20}.

Throughout the paper, we use units where $\hbar=c=1$.


\section{Kinematics of Twisted-Photon Absorption}

\subsection{Angular-Momentum Conservation and a Superkick}

\begin{figure}[h]
\centering
\includegraphics[height = 75 mm]{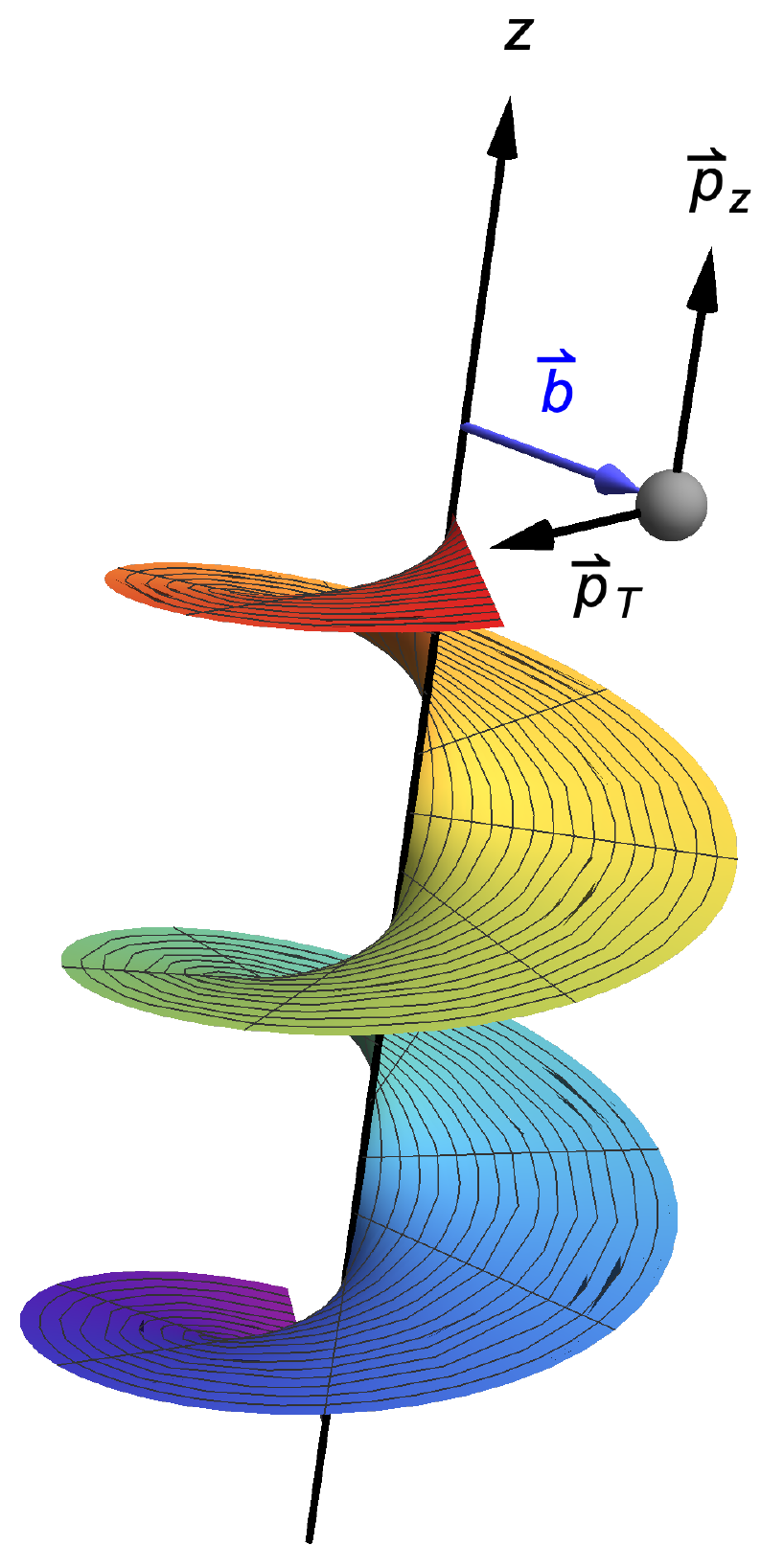} \hfil

\caption{Twisted photon's helical wavefront and an atomic target located at an impact parameter $b$ from the photon's axis $z$ (or phase singularity). The momenta $p_T$ and $p_z$ show transverse recoil and longitudinal recoil, respectively.}
\label{fig:Wave}
\end{figure}

Let us consider an atom, or another sub-wavelength-size target that absorbs a twisted photon; the target is located at a distance $b$ away from the photon's axis, as shown in Figure \ref{fig:Wave}. The formalism for calculating individual quantum transition amplitudes due to absorption of the twisted photons can be found elsewhere, \cite{Afanasev:2013kaa,2014PhRvA..90a3425S,Afanasev_2016}; here we are concerned with a magnitude of recoil momentum $p_T$ of the target after photo-absorption.

The transverse kick given to a target atom offset distance $b$ from the vortex line of the photon relates directly to the AM transferred to the atom's overall center of mass.   Hence we start by  considering the average angular momenta given to the internal electronic state and to the atomic c.m.~in a photoexcitation process.

The expectation value $\langle \ell_z \rangle$ of AM  transferred by a twisted photon with AM $z$-projection $m_\gamma$ to internal degrees of freedom of an atom can be expressed in terms of the probabilities $w(m_f)$ for exciting the atom to states with magnetic quantum numbers $m_f$~\cite{curvenote,Afanasev_2020},
\begin{equation}
\label{eq:Lz}
\langle \ell_z \rangle =\sum_{-\ell<m_f<\ell} m_f \, w(m_f).
\end{equation}
From AM conservation, the initial AM not transferred to the internal excitation goes to the atom's c.m.~motion~\cite{Afanasev:14}, 
\begin{equation}
\label{eq:LzAtom}
\langle \ell_z \rangle_{c.m.} =m_{\gamma}-\braket{l_z}.
\end{equation}

\begin{figure}[h]
\centering
(a) \\
\includegraphics[width = 0.85 \columnwidth]{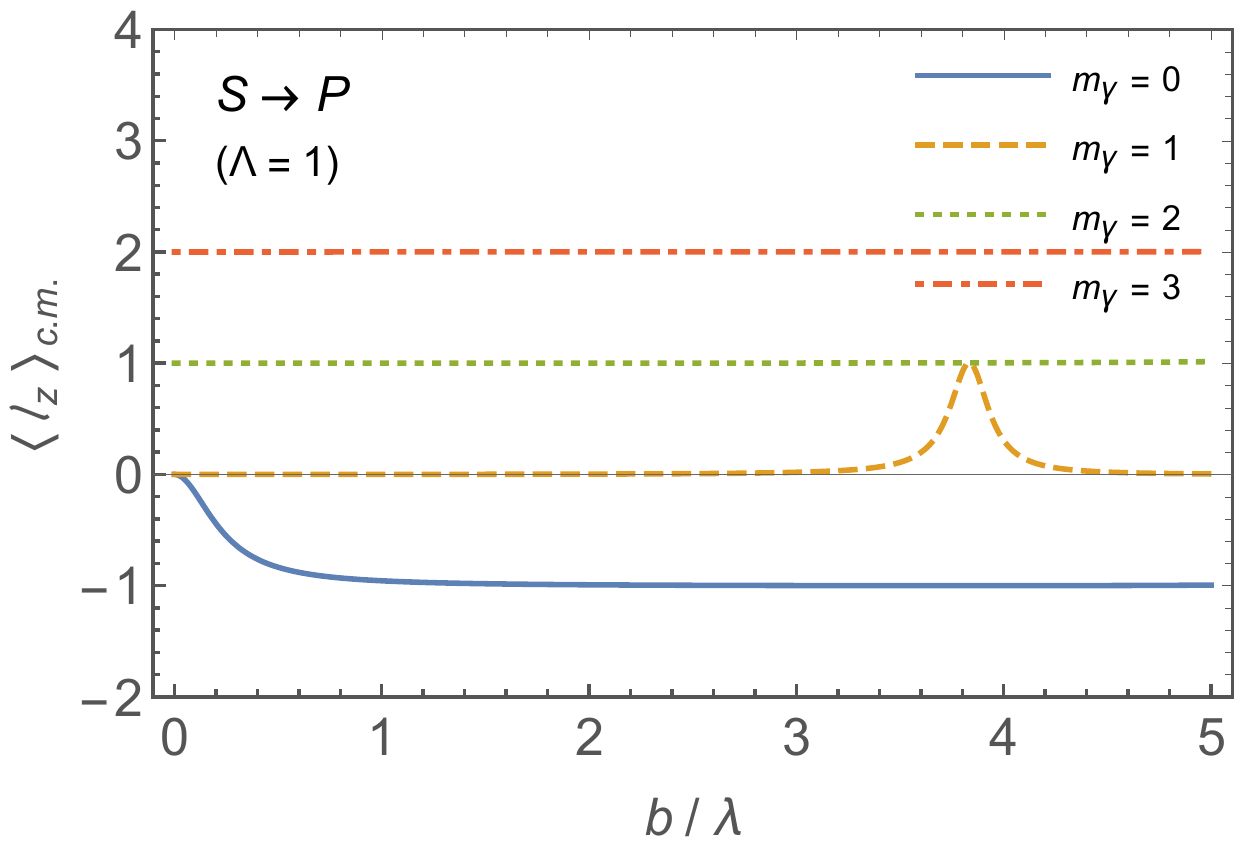} \\[2 ex]
(b)	\\
\includegraphics[width = 0.85 \columnwidth]{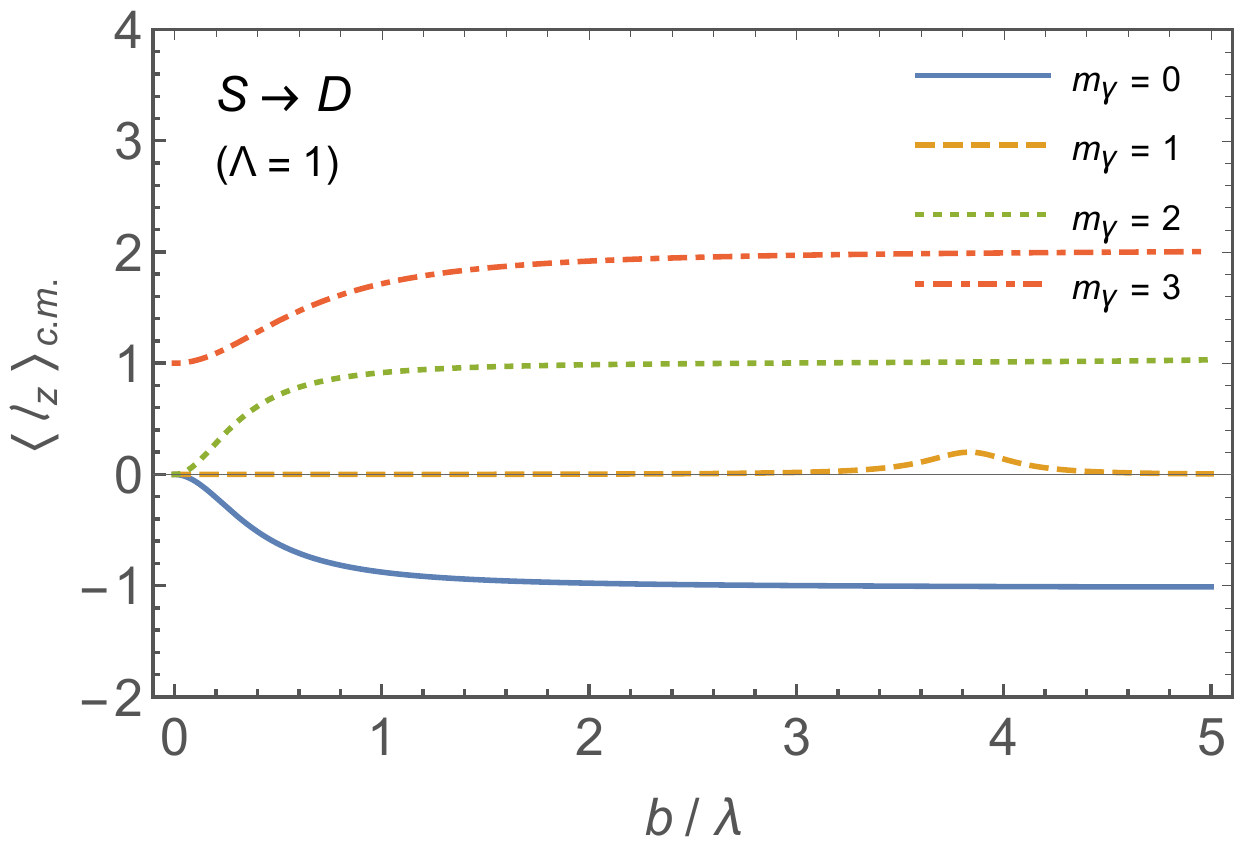}	\\[2 ex]
(c)	\\
\includegraphics[width = 0.85 \columnwidth]{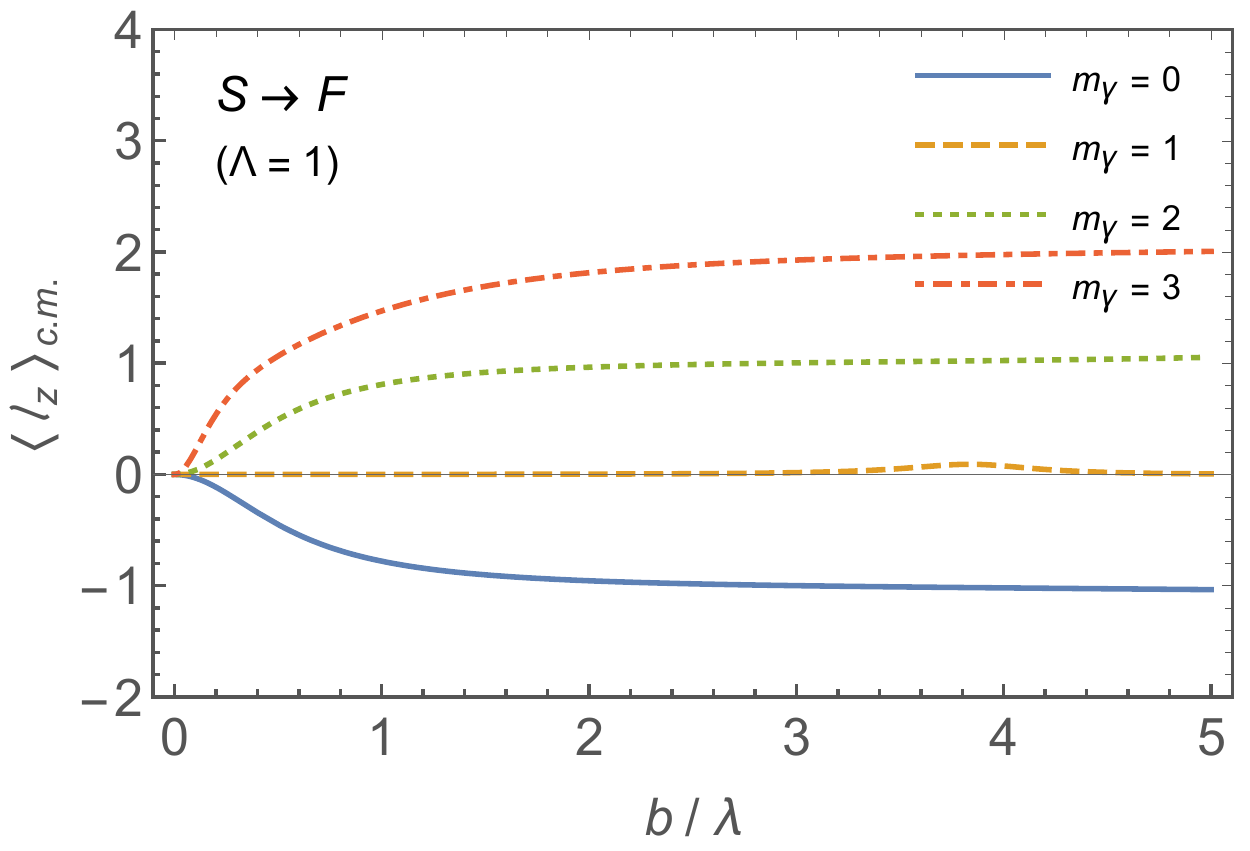}

\caption{Mean angular momentum transfer $\langle \ell_z \rangle_{c.m.}$, Eq.~(\ref{eq:Lz}), along the beam direction passed by twisted light of total angular momentum $m_\gamma$ to an atom's c.m. motion for (a) $S\to P$ transitions, (b) $S\to D$ transitions, and (c) $S\to F$ transitions. For all cases, $\Lambda = +1$ where $\Lambda$ is (paraxially) the spin angular momentum of the twisted photon. The horizontal axis shows atom's position $b$ with respect to the vortex center measured in units of light's wavelength. See text for further comments. }
\label{fig:Lz1}
\end{figure}

\begin{figure}[h]
\centering
(a) \\
\includegraphics[width = 0.85 \columnwidth]{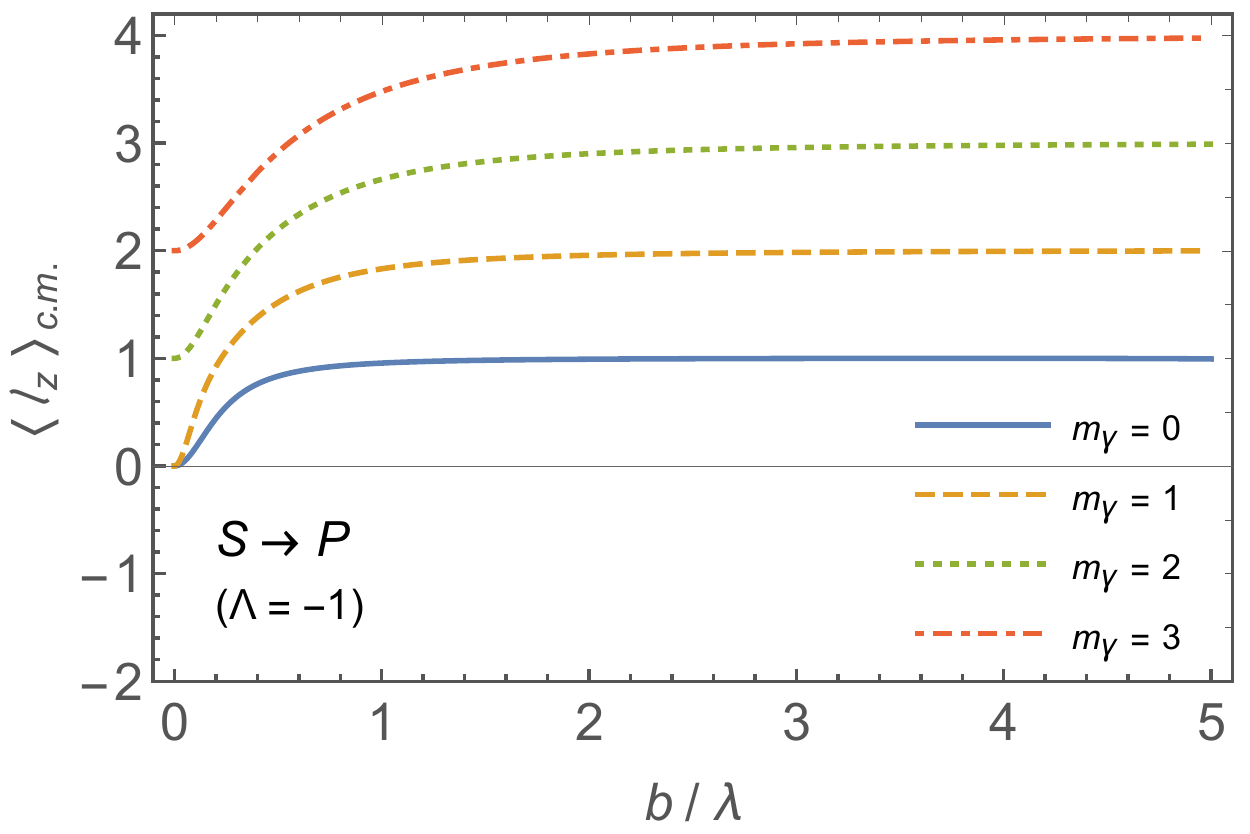} \\[2 ex]
(b)	\\
\includegraphics[width = 0.85 \columnwidth]{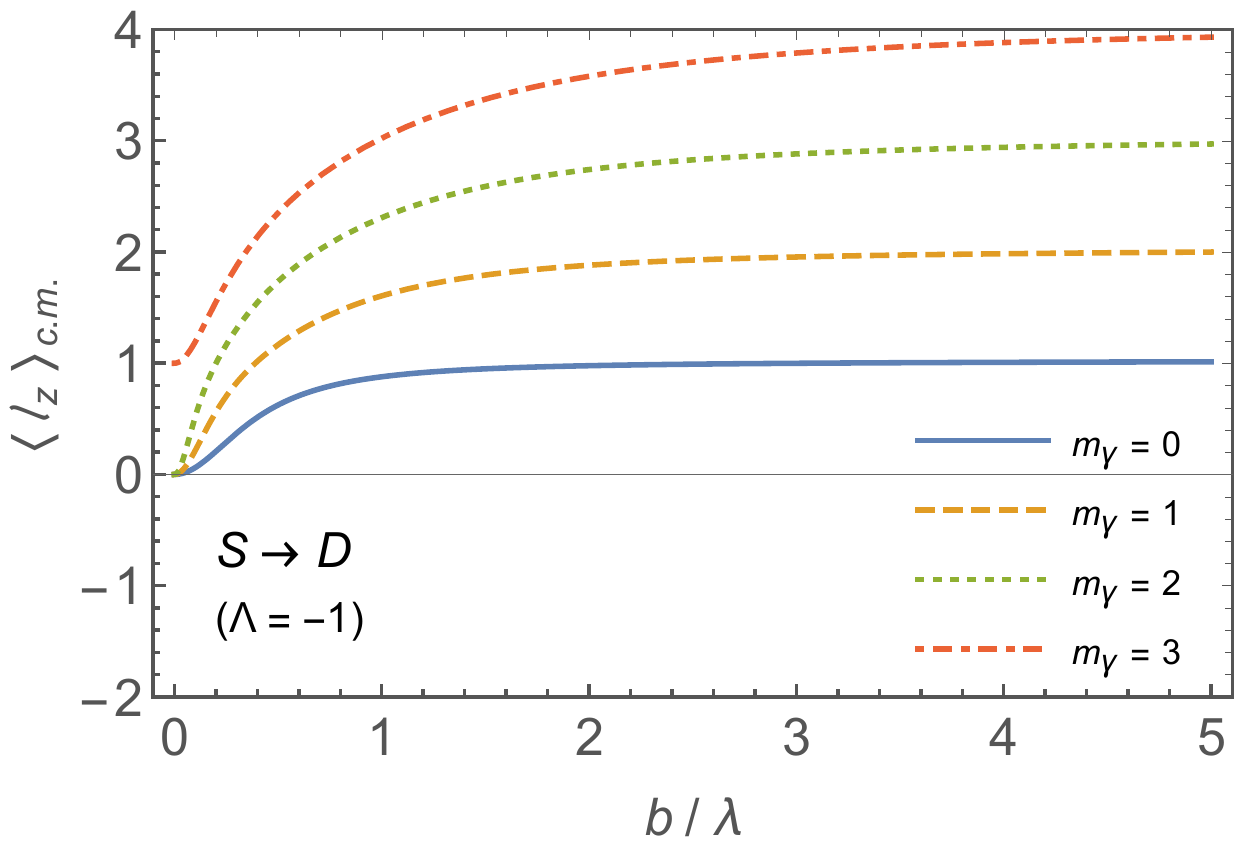}	\\[2 ex]
(c)	\\
\includegraphics[width = 0.85 \columnwidth]{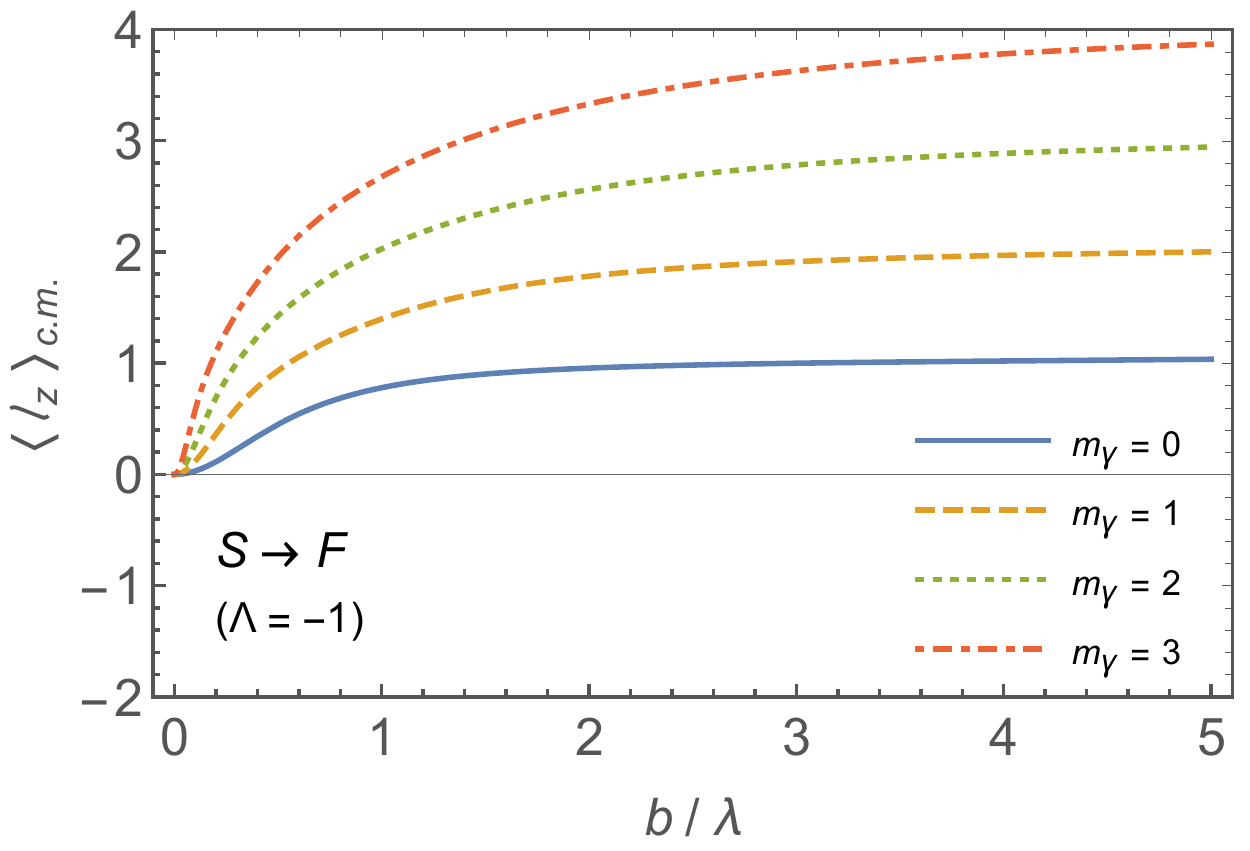}

\caption{Same as Fig.~\ref{fig:Lz1} but for $\Lambda = -1$. }
\label{fig:Lz-1}
\end{figure}

Plots of AM transfer vs impact parameter $b$ are shown in Figs.~\ref{fig:Lz1} and~\ref{fig:Lz-1}. We considered $S \to P$, $S \to D$, and $S \to F$ atomic transitions for several choices of incoming twisted photon quantum numbers as labeled in the Figures.  With exceptions at some values of $b$, for $S \to P$ transitions the atoms still absorb just one unit of angular momentum into their electronic degrees of freedom, just like for plane waves (with $\Delta m=\pm1$ dipole selection rules), leaving the rest of the AM for c.m. motion.  For $S \to D$ and $S \to F$ transitions, the AM transfers to $\braket{l_z}_{c.m.}$  deviate from plane-wave selection rules for smaller, sub-wavelength, values of $b$,  especially when the total incoming photon AM is greater than a single  $\hbar$. The calculation applies for the case when magnetic quantum numbers of the excited atom are not resolved. There is also a possibility to measure individual transitions into Zeeman sublevels if these levels are split by an external magnetic field, as was done in Refs.\cite{2016NatCo...712998S,Afanasev_2018}.

While the longitudinal momentum of the atom's recoil equals photon's longitudinal momentum $p_z=\hbar \omega /c$ (in the paraxial approximation, and where $\omega$ is photon's angular frequency), the transverse recoil momentum $p_T$ can be evaluated through AM conservation as $p_T=\langle \ell_z \rangle_{c.m.}/b$ (at least for values of $b$ well larger than the radius of the target~\cite{Barnett_2013}).  Therefore their ratio is:
\begin{equation}
\frac{p_T}{p_z}= \frac{\langle \ell_z\rangle_{c.m.}}{2\pi \hbar} \frac{\lambda}{b} \,,
\end{equation}
where $\lambda$ is twisted photon's wave length.

For the case when photon's spin ($\Lambda$) and total AM ($m_\gamma$) are aligned, and the atomic transition is $S\to P$, as shown in Fig.\ref{fig:Lz1}a, the orbital AM   is given by $\langle \ell_z\rangle_{c.m.}=\hbar(m_\gamma-\Lambda)$ and we obtain a simple formula for the transverse recoil momentum,
\begin{equation}
p_T=\hbar\frac{m_\gamma-\Lambda}{b}
\label{eq:pT}
\end{equation}
It follows that the longitudinal and transverse recoil momenta become equal at the value of impact parameter
\begin{equation}
b=\lambda \frac{m_\gamma-\Lambda}{2 \pi}.
\end{equation}

\begin{figure}[h]
\centering
(a) \\
\includegraphics[width = 0.85 \columnwidth]{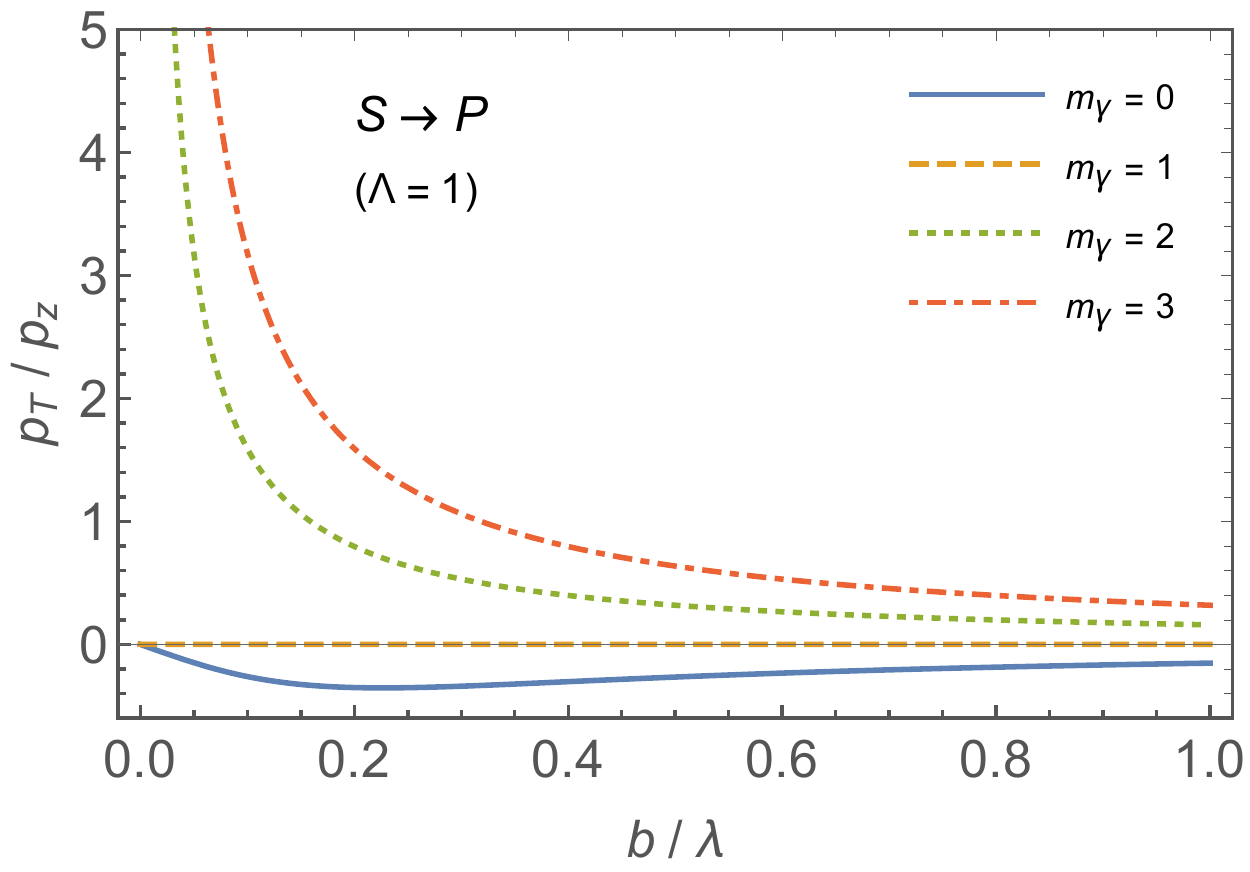} \\[2 ex]
(b)	\\
\includegraphics[width = 0.85 \columnwidth]{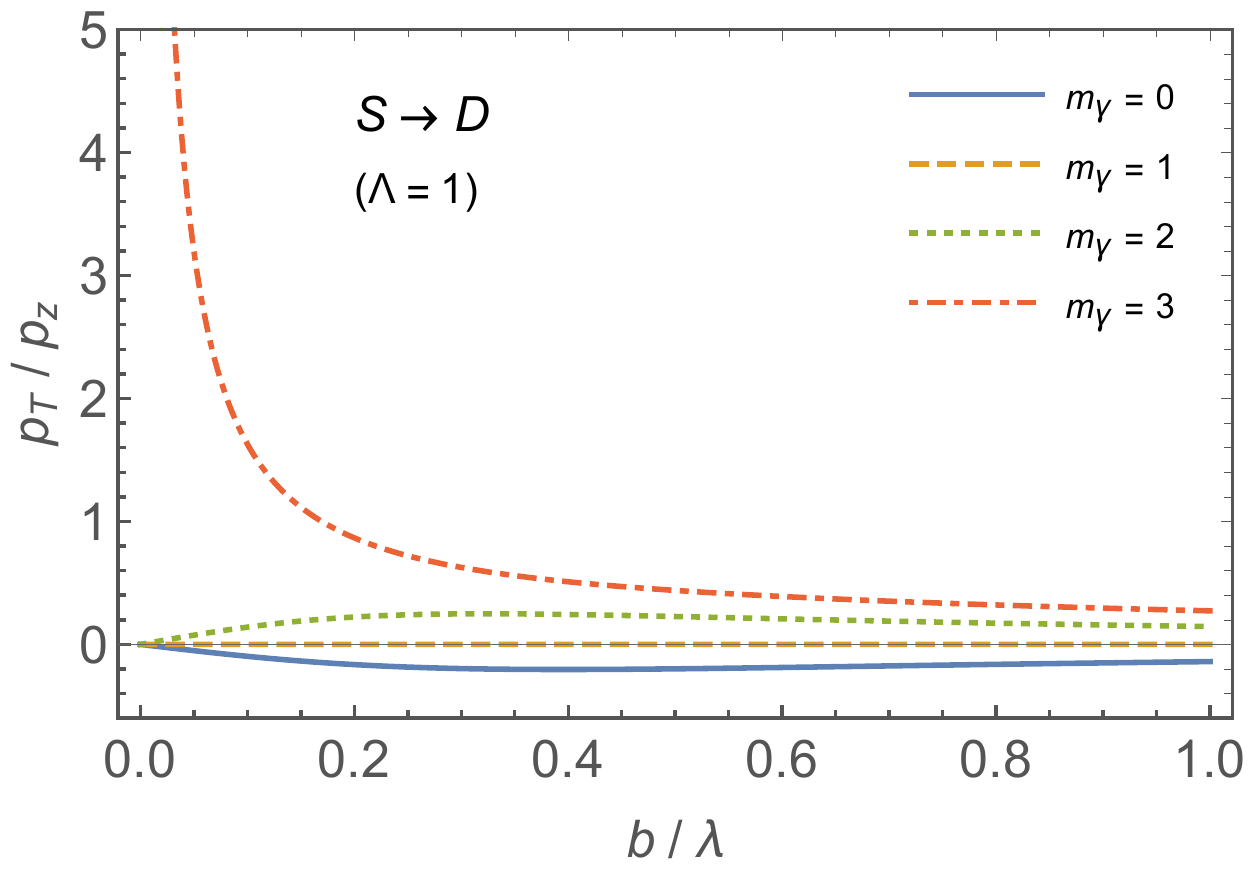} \\[2 ex]
(c)	\\
\includegraphics[width = 0.85 \columnwidth]{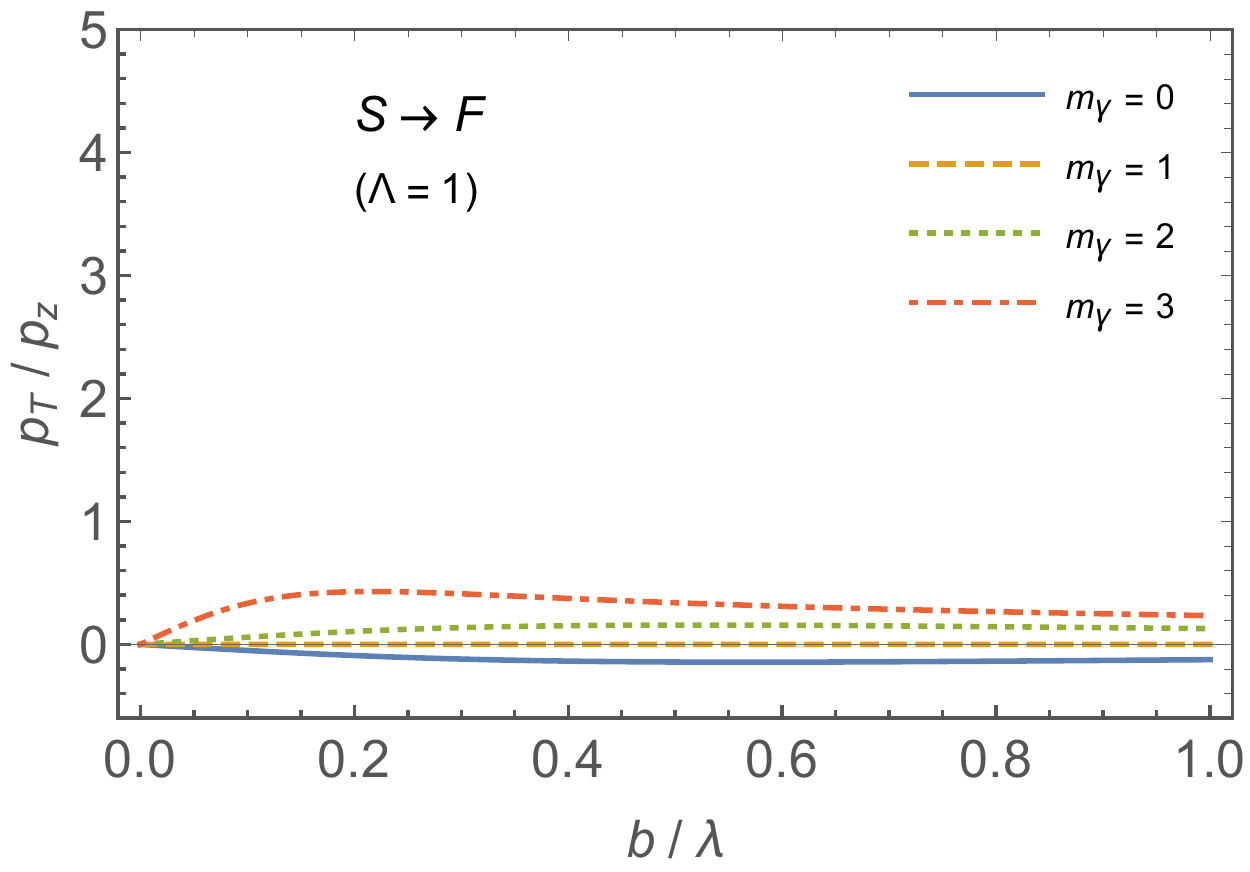}

\caption{Ratio of longitudinal to transverse recoil momentum of a free atomic target after absorbing a twisted photon in  $S\to P$ transition (a) $S\to D$ transition (b) and $S\to F$ transition. For all cases, $\Lambda = +1$, $i.e.$ the spin of a twisted photon is aligned with its orbital AM. The horizontal axis shows atom's position $b$. }
\label{fig:pRat1}
\end{figure}

\begin{figure}[h]
\centering
(a) \\
\includegraphics[width = 0.85 \columnwidth]{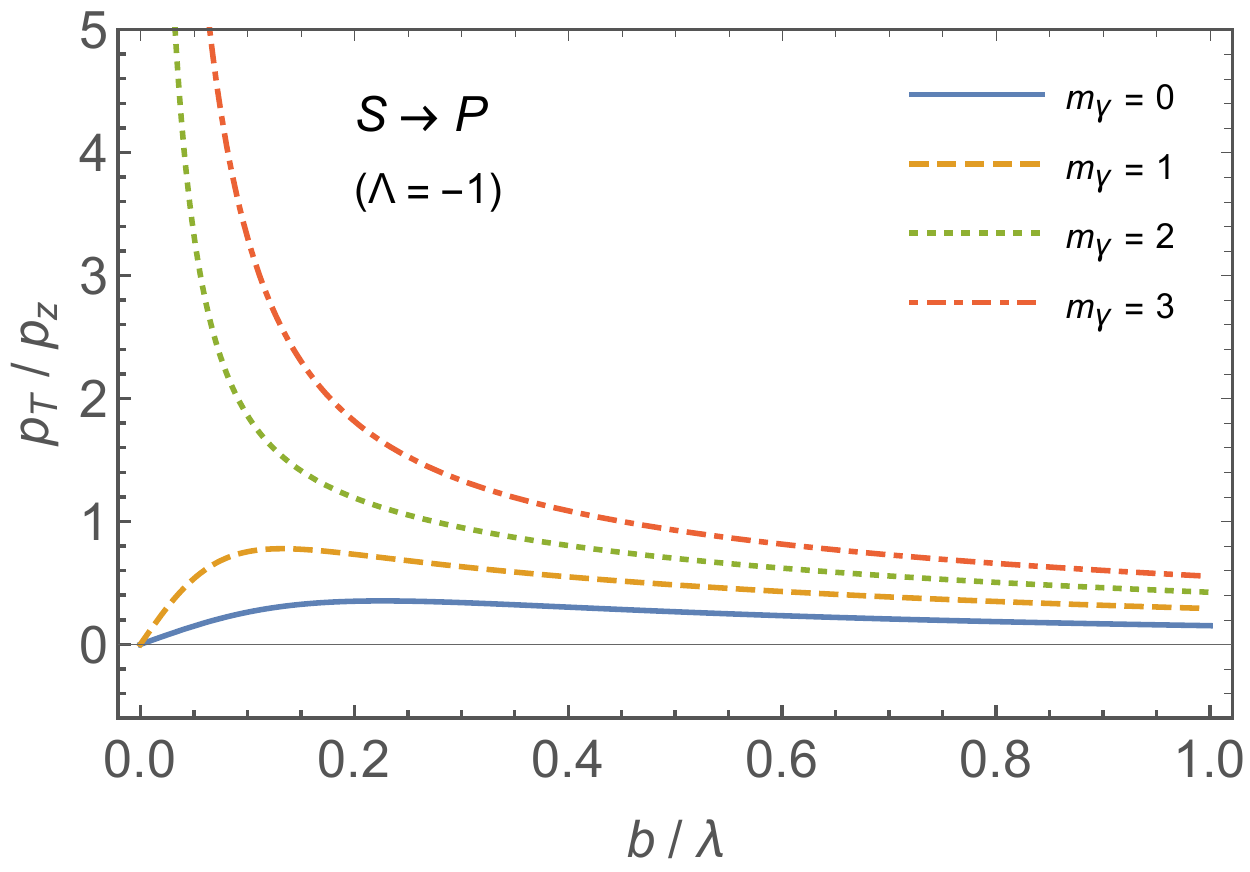} \\[2 ex]
(b)	\\
\includegraphics[width = 0.85 \columnwidth]{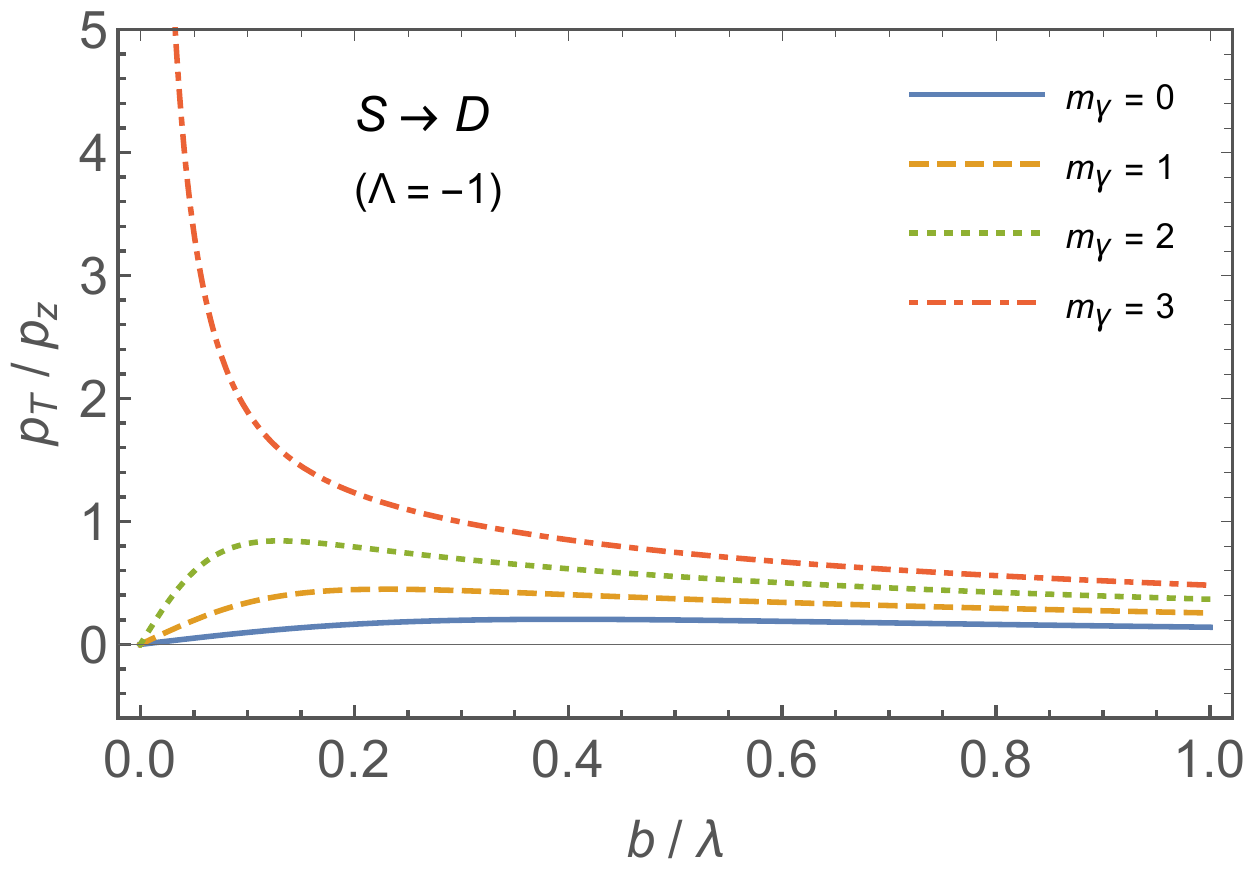} \\[2 ex]
(c)	\\
\includegraphics[width = 0.85 \columnwidth]{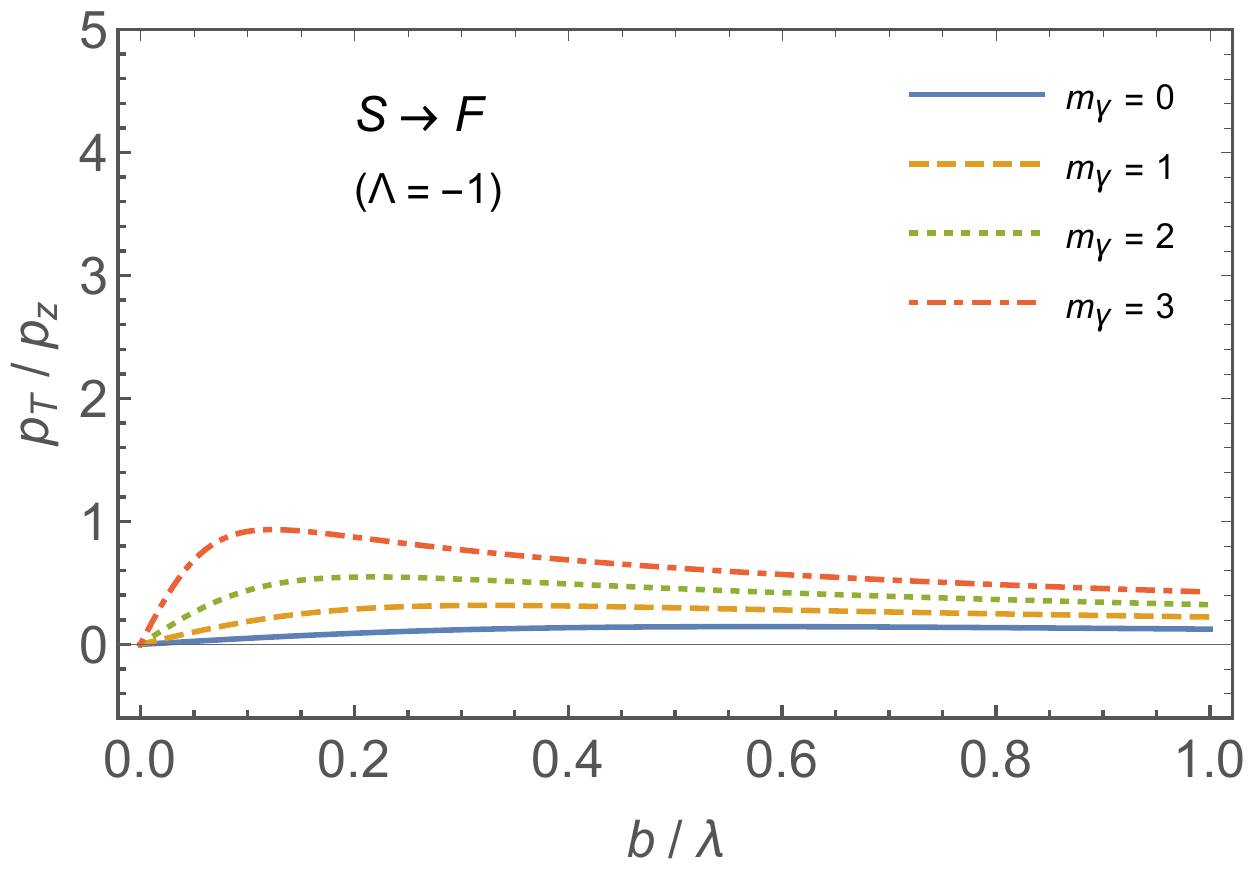}

\caption{Same as Figure ~\ref{fig:pRat1} but for $\Lambda = -1$. }
\label{fig:pRat-1}
\end{figure}

However, if some of the excess AM of the incoming twisted photon is passed to internal excitation of the target, then the c.m. recoil is dampened at sub-wavelength distances near the vortex center. This effect is shown for the ratios $p_T/p_z$ in Figs.~\ref{fig:pRat1},~\ref{fig:pRat-1}. Qualitatively, this effect was discussed in Ref.~\cite{Babiker_2018}, but specific predictions for non-dipole transitions are presented here for the first time, to the best of our knowledge.

The above results indicate that for $S\to P$ transitions (see Fig.~\ref{fig:Lz1}a and \ref{fig:pRat1}a for $m_\gamma=2,3$), the approach of Barnett and Berry \cite{Barnett_2013}  to the evaluation of atomic recoil for absorption of the twisted light is justified. They do have general 
$m_\gamma$, but limit consideration to a single level final state and only have dipole transitions.     For general cases, modification is required, as shown here. 


\subsection{Twisted Photon Absorption on Cold Trapped Ions}


Let us consider atomic recoil of a $^{40}$Ca$^+$ ion after absorption of 397 nm photon in an $S\to P$ transition or 729 nm photon in an $S\to D$ transition that define the ``carrier" frequency of the absorbed photons.

In presence of atomic target recoil, energy conservation is modified as follows,
\begin{equation}
\hbar \omega=\hbar \omega_0+\frac{p_z^2+p_T^2}{2M},
\label{eq:E_th}
\end{equation}
where $\hbar \omega_0$ define the excited energy level. 

\begin{figure}[t]
\centering
\includegraphics[width = 0.98 \columnwidth]{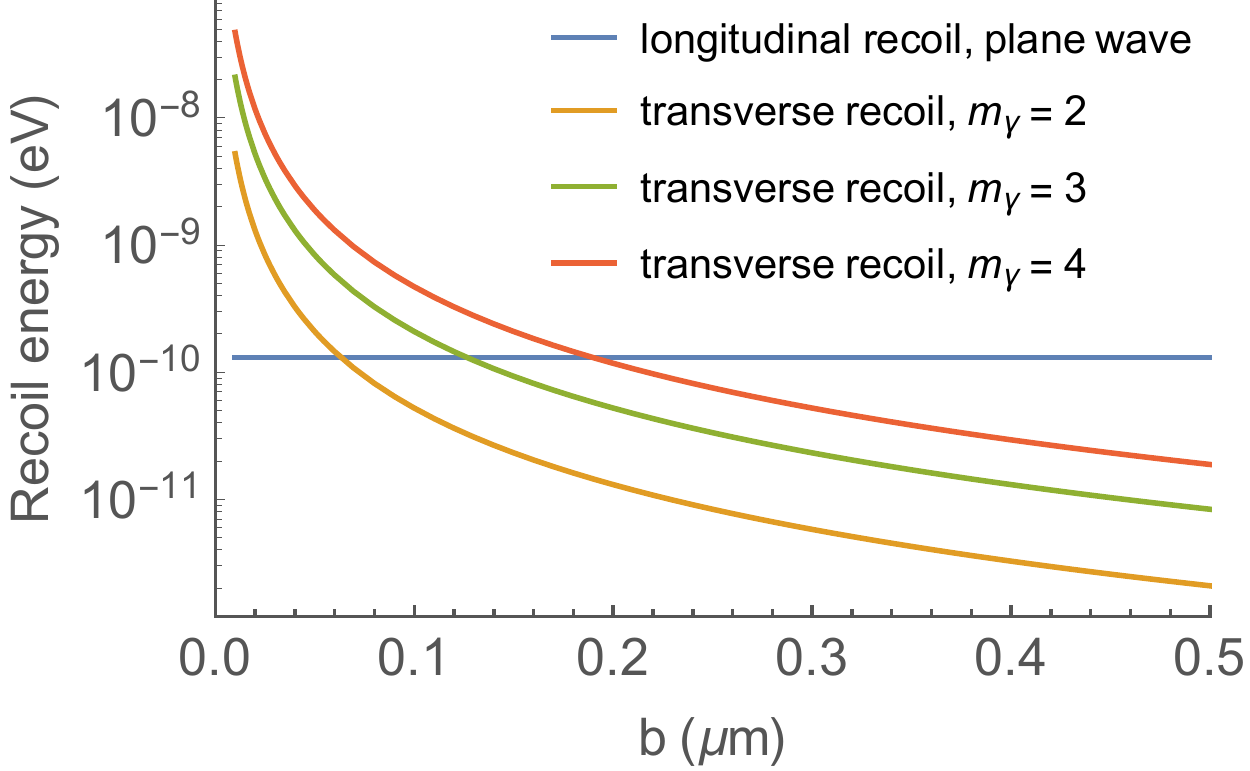} \hfill

\caption{Recoil energy as a function of impact parameter for longitudinal recoil and transverse recoil at different values of total AM of the absorbed photon, $\Lambda=1$, for  $\lambda$ = 397 nm $E1$ $S\to P$ transition on $^{40}$Ca$^+$ ion.}
\label{fig:40Ca}
\end{figure}

If, for example, a free ion of $^{40}$Ca absorbs a plane-wave photon of wavelength $\lambda=397$ nm and energy E$_\gamma = 3.12$ eV, corresponding to an $E1$ $S\to P$ transition, it gives an atom longitudinal recoil energy of $p_z^2/(2M)=0.13$ neV, where $M$ is target's mass. Twisted-photon absorption generates additional transverse recoil momentum that depends on the impact parameter $b$ but is independent on photon's wavelength. It can be read off Figs.~\ref{fig:pRat1}a and~\ref{fig:pRat-1}a.  The conversion into transverse recoil energy, $E_T=p_T^2/(2M)$, is plotted in Fig.~\ref{fig:40Ca}, with a comparison line for the longitudinal recoil energy $p_z^2/(2M)$.   For the $S \to D$ electric quadrupole $E2$ transition at $\lambda = 729$ nm, the corresponding values of transverse recoil momentum can be read off Figs~\ref{fig:pRat1}b and~\ref{fig:pRat-1}b.

In actual experiments the ions are being held in electromagnetic traps; for example, a segmented Paul trap was used in Refs.~\cite{2016NatCo...712998S, Afanasev_2018} with RF frequencies of about $f=\omega^z_\text{trap} /2\pi=1.5$ MHz (along the trap's axis $z$) which corresponds to 6.2 neV energy level spacing in a harmonic oscillator. Different frequencies 
$\omega^{x,y}_\text{trap}$ describe transverse motion of ions in the trap. From the above, we can estimate the value of impact parameter $b=10$ nm for which  the transverse recoil equals the energy level spacing of the trap.  It affects the Lamb-Dicke parameter $\eta$ that is crucial for  determining ion behavior in the trap. 
It can be obtained from $\eta=\sqrt{E_\text{rec}/\hbar \omega_\text{trap}}$, where $E_\text{rec}$ is the recoil energy.   The condition $\eta \ll 1$ defines a Lamb-Dicke regime important for cooling the ions down to the oscillator ground energy of the trap; see Ref. \cite{Eschner:03} for details. It implies that the superkick generating $E_\text{rec}\approx\hbar \omega_\text{trap}$ results in breaking of  a Lamb-Dicke regime for transverse ion motion at sufficiently small values of the impact parameter, depending on light's orbital AM and trap's frequency.   

\begin{figure}[t]
\begin{center}
\includegraphics[width = 0.98 \columnwidth]{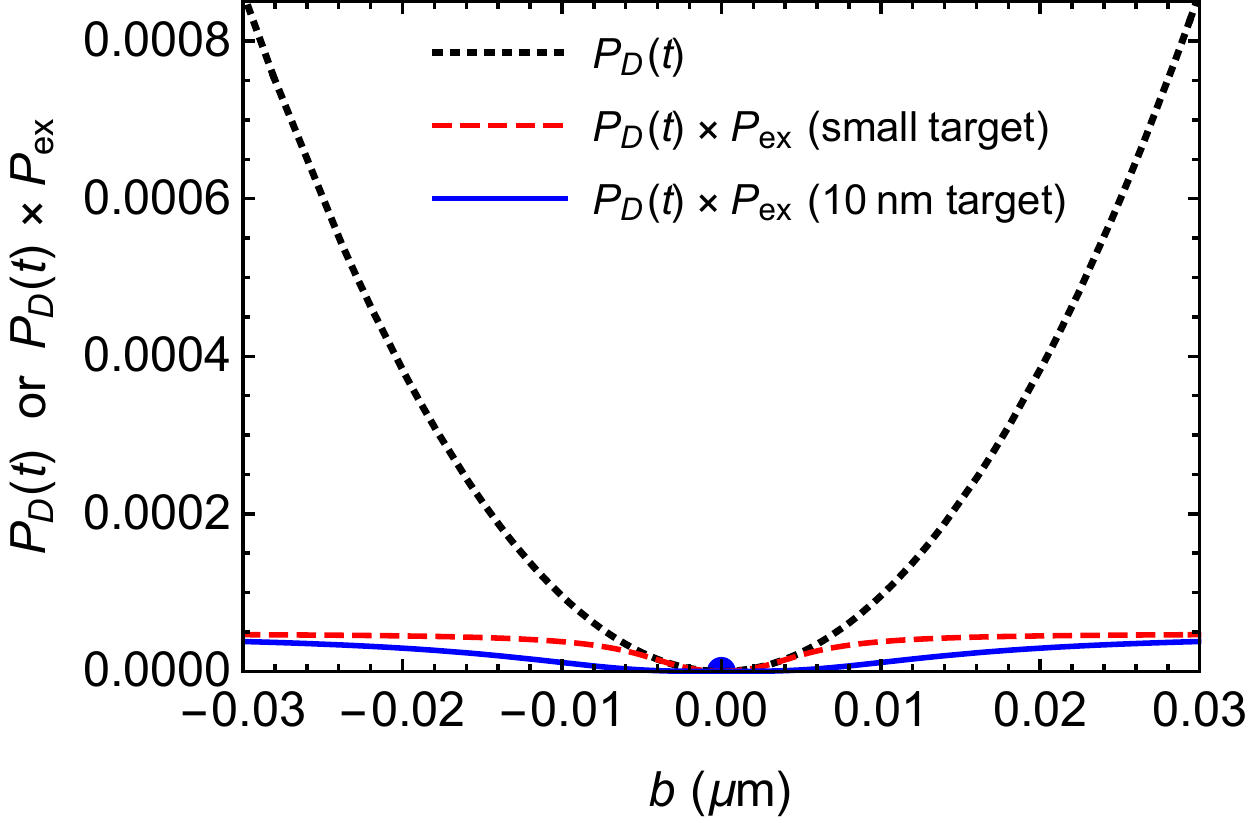}
\caption{The black dotted curve shows the probability for an $S_{1/2} \to D_{5/2}$ transition, specifically for $m_\gamma = -2$,  $m_i = -1/2$, $m_f = -3/2$, and a measurement time of $26\,\mu$s (the same as for the data for the same transition in~\cite{Afanasev_2018}).   Also shown is the transition probability multiplied by the further probability that the atom is put into an excited state in its trapping potential by the linear transverse kick.  The latter is shown for two cases:  a red dashed line for an atom with an almost pointlike spatial distribution, and  a blue solid line for a realistic case in which an atom is spread over some region in a harmonic oscillator potential, in this case, with a 10 nm rms spread.}
\label{fig:jumpfraction}
\end{center}
\end{figure}

A  consequence of the larger Lamb-Dicke parameter, or of the recoil energy, is to move the atom into a higher state of the confining potential.  Some estimate of how often this happens is shown in Fig.~\ref{fig:jumpfraction}.  The transition probability is shown with a black dotted line for a particular situation involving twisted photons (in this case $S \to D$ transitions in $^{40}$Ca$^+$ with $m_\gamma = -2$, the initial state with electron spin projection $-1/2$, the final state having magnetic quantum number $m_f = -3/2$, and a measurement time of $26\,\mu$s~\cite{Afanasev_2018}).   

Also shown Fig.~\ref{fig:jumpfraction} is the $S \to D$ transition probability multiplied by the further probability that the ion will jump to an excited state of the trapping potential if the atom is confined to a very small region (labelled ``point target''), and the similar probability if the atom is confined to a finite size region.  In the latter case,  the superkick is reduced because the atom's wave function in the potential  can saddle the vortex line.       For the $S \to D$ transition probability, the calculation was done using methods detailed in~\cite{Afanasev_2018},  and the excitation probability was modeled using Gaussian wave functions for the trapped atom, with results similar to~\cite{Barnett_2013}.

The calculation and  Fig.~\ref{fig:jumpfraction} show that there is a jump
to an excited state for about one transition in 15 due to the linear superkick, which
could be enough to see, if one can do measurements at
small impact parameters. It should be noted that the
shift in absorption energy is a position-dependent combined effect of c.m.~motion of the oscillator and internal rotation of the oscillator. If the twisted-photon beam
is aligned along the trap's $z$-axis, and the trap's confining
potential is isotropic in the $x, y$-directions ($\omega_\text{trap}^{x,y} = \omega^T_\text{trap}$),
then the superkick effect will result in emerging sidebands
at energies shifted by $\pm \omega_\text{trap}^T$.  The relative strength of these sidebands with respect to the main ``carrier'' frequency of the atomic transition depends on the impact parameter $b$, the width of trapped ion's wave packet, and whether or not the transverse motion is in the ground state (for which red-shifted sideband $- \omega_\text{trap}^T$ is not allowed).
For the beam centered on the trapped ion's wave packet, the twisted-photon absorption results in either red or blue shift of absorbed photons, and disappearance of the ``carrier" frequency absorption line.


We add a few comments.  Ion cooling in optical molasses produced by twisted light for the purposes of quantum computing and/or precision fundamental measurements was discussed in the literature, with a proposal in Ref.\cite{Lembessis_2011} to use polarization gradients formed in the superposition of the twisted light beams. For more discussions and approaches, see review \cite{Babiker_2018}. Here, we emphasize that for ions located at $b\leq \lambda(m_\gamma-\Lambda)/2\pi$ the transverse recoil in electric-dipole single-photon absorption may exceed longitudinal recoil, and this fact has to be taken into account in possible ion-cooling scenarios. 

Due to increased transverse recoil of the trapped ions, the corresponding Lamb-Dicke parameters will be increased in the vicinity of the optical vortex center, that may affect both Doppler cooling and sideband cooling processes \cite{Eschner:03} required for quantum computing and precision measurements. Experimental observation of the change in laser cooling dynamics of trapped ions can potentially provide a method to study the superkick effect.

Finally, we point out the fact that the transverse recoil passed to the trapped ions can generate quantized 3D motion characterized by discrete values of orbital AM. It may offer new opportunities for quantum state manipulation in quantum computing applications.

\subsection{Photo-Disintegration of Deuterium}


Let us now consider one of the simplest photo-nuclear processes, photo-disintegration of a deuteron
$$
\gamma+D\to n+p,
$$
where the symbols $D$, $n$, and $p$ stand for deuteron, neutron, and proton, respectively. The deutron's binding energy is known with sub-keV accuracy, $E_B=2224.52 \pm 0.2$ keV.
Since the binding energy is much smaller than deuteron's mass, we can use Eq.~(\ref{eq:E_th}) to evaluate the reaction threshold. Like in the case of ion excitation, in the dipole ($M1,E1$) transitions that dominate near the threshold~\cite{BlattWeisskopf}, the transverse recoil momentum depends on deuteron's position and increases toward smaller values of impact parameter $b$. For plane-wave incident photons with wavelength $\lambda=559$fm, the recoil energy at the threshold is $E_B^2/2Mc^2=1.3$ keV, which is larger than the current $0.2$ keV error in deuteron's binding energy. For the twisted photons providing one extra unit of AM to c.m. motion of the final (np)-state, the recoil energy would double to 2.6 keV for $b= \lambda/2\pi=89$ fm and would increase as $1/b^2$ for smaller values of $b$ until the uncertainty in deuteron's spatial location becomes a limiting factor. 
The probability of this reaction due to twisted photons in the dipole approximation was considered in Ref.~\cite{AfanasevSerbo_2018}; we point out here that for the same approximation the reaction threshold is increased as a function of nucleus's position with respect to the vortex axis of a twisted photon.

Would it mean that the reaction threshold is modified overall for the above reaction? The answer is `not always', because another allowed transition near the threshold is electric quadrupole ($E2$) into the triplet $^3S_1$ state of unbound neutron and proton, which is due to a relatively small, $\approx5$\% admixture of $D$-state in the deuteron's wave function. If the twisted photon's total AM equals $m_\gamma=2\hbar$, in  $E2$-transition the target would absorb this AM into its intrinsic degrees of freedom, and not in c.m. motion, the extra recoil would be the same as in the plane-wave photon case and the threshold for $E2$ transition remains unchanged in the vicinity of the vortex center. However, for higher values of $m_\gamma$, the threshold energy must also be increased, although with a larger phase space for quadrupole transitions than for dipole ones.  Independently, the matrix elements of quadrupole transitions can also be enhanced compared to dipole due to novel AM selection rules in the twisted-photon absorption, see Ref.~\cite{Afanasev_2018_spectroscopy,Solyanik-Gorgone_2019}.

A possible way to observe the superkick in this photonuclear process would be through analysis of the energy spectrum of final protons and/or neutrons. For kinematics near the reaction threshold, the (np)-pair is produced with a small relative momentum and moves almost parallel to the incident photon. Since the additional recoil momentum from twisted $\gamma$-rays is purely transverse, measuring energies of either of the nucleons emerging at large angles (with respect to the beam) and having an energy excess would indicate the superkick effect; the excess recoil energy can be measured by proton or neutron spectrometry or time-of-flight methods. If observed, these nucleons with $\ge$ keV energy excess would pinpoint the spatial location of the reaction to a hundred of femtometers within the phase singularity of the twisted $\gamma$-ray. 

Tight focussing of twisted gamma-ray beams is essential for feasibility of such measurements. We estimate that if the beam is focussed to about 50 picometers, then approximately one per cent of final nucleons would have transverse momenta exceeding one-tenth of their longitudinal momentum. If, however, the beam is focussed within 3 picometers, then the number of such nucleons reaches 99 per cent. 

The above findings imply that for a range of twisted-photon energies near the threshold of deuteron photo-disintegration, one can observe a complete absence of $E1$ or $M1$-transitions, while the only surviving transitions will be electric quadrupole $E2$, sensitive to deuteron's $D$-state admixture  known to be a fundamental property of nucleon-nucleon interactions in need to precision determination. Thus, twisted photons may become a new tool for nuclear physics studies; a possibility of their generation for nuclear studies via Compton backscattering, or inverse Thompson scattering, was discussed in Refs.~\cite{Jentschura_2011,Petrillo_2016}.

\subsection{Electron-Positron Pair Production by High-Energy Cosmic Rays}

Previous examples were concerned with heavy targets compared to the absorbed photon energy, with recoil energies in the range of neV for atomic processes and keV for nuclear processes. Let us now consider a target of zero mass, namely, a photon. In particular, we are interested in collisions of VHE $\gamma$-rays with optical-energy photons leading to creation of electron-positron pairs. 

Our motivation for this section comes from still unexplained transparency of Universe to VHE cosmic rays in multi-TeV energy range that are expected to be absorbed on extra-galactic background light (EBL)  via electron-positron pair production process $\gamma\gamma\to e^+e^-$, but measured spectra of VHE gammas originating from remote blazars did not show noticeable signs of attenuation  \cite{Aharonian_2006}. These observations were interpreted as indication of physics beyond Standard Model (BSM) \cite{deAngelis_2007}, and are presently considered as an indirect evidence of dark matter \cite{Jaeckel_2010,Anantua:2010zz}. Further measurements \cite{Albert_2008} confirmed that the cosmic ray spectrum does not show onset of attenuation toward higher energies expected from pair production; see also Ref.~\cite{Madejski_2016} for the most recent updates. 

In the reaction
\begin{equation}
\gamma(k_1)+\gamma(k_2)\to e^-(p_1)+e^+(p_2),
\end{equation}
where each particle is labeled with its corresponding 4-momentum.   Let us assume that one of the photons is VHE ($k_1$), coming from an unspecified astrophysical source, while another photon is optical ($k_1$), due to the EBL.

For reference, if both photons be plane waves, the energetic photon energy must satisfy
\be
\omega_1 \ge \frac{ m_e^2 }{ \omega_2 } \approx 100 \text{\,GeV}		,
\ee
where $\omega_2$ is the energy of the extragalactic background photon and the numerical value is for $\omega_2 = 2.5$ eV, a typical number for green light.

For the twisted photon, described theoretically by a Bessel beam, the Fourier representation in wavenumber space consists of photons all with the same energy, same $k_z$, and all moving at the same polar angle or pitch angle to the $z$-axis, with varying azimuthal angles.  The energy and longitudinal momentum are then specified by
\be
k_1 = ( \omega_1, \,.\, , \,.\, , \omega_1 \cos\theta_k )	,
\ee
where $\theta_k$ is the pitch angle, or the conical opening angle of plane-waves that form Bessel wave packet, see e.g., Ref.\cite{Afanasev:2013kaa} for definitions.  The transverse momentum depends on the location within the wave front in coordinate space and will be discussed momentarily.   At extreme energies the slight difference of the pitch angle from zero enters in the longitudinal momentum and will matter.  The reduction in $p_z$ by the $\cos\theta_k$ factor means a smaller longitudinal kick to the final state, reducing the kinetic energy need,  suggesting a reduction in the threshold energy, at least in the absence of a transverse kick.    

The transverse kick will give an effect in the opposite direction.  The transverse kick occurs because the twisted wave front is swirling and the local transverse momentum of the state is, in magnitude, 
$p_T = \frac{ \ell_\gamma \hbar }{ b } = \frac{ (m_\gamma - \Lambda) \hbar }{  b}$, as given by Eq.(\ref{eq:pT}). A final state produced at a distance $b$ from the vortex line receives a transverse momentum kick of this magnitude.   

If the final state is produced at a definite location distance $b$ away from the vortex line, the $e^+ e^-$ total momentum will be
\be
P = \left( \omega_1+\omega_2 , p_T , \omega_1 \cos\theta_k - \omega_2 \right)
\ee
for a head on collision, with $p_T$ given above.  This leads to the threshold relation
\be
\label{eq:general}
\omega_1^2 \sin^2 \theta_k + 4 \omega_1 \omega_2 \ge 4 m_e^2 + p_T^2	\,.
\ee
If $\theta_k$ is very small,
\be
\omega_1 \ge \frac{m_e^2}{ \omega_2 } + \frac{ p_T^2 }{ 4 \omega_2 }	\,.
\label{eq:ggRecoil}
\ee
The first term echoes the plane wave result, and the second term increases the required threshold energy in response to the additional momentum in the transverse direction.  

\begin{figure}[t]
\begin{center}
\centerline{(a)}

\centerline{  \includegraphics[width = 0.92\columnwidth]{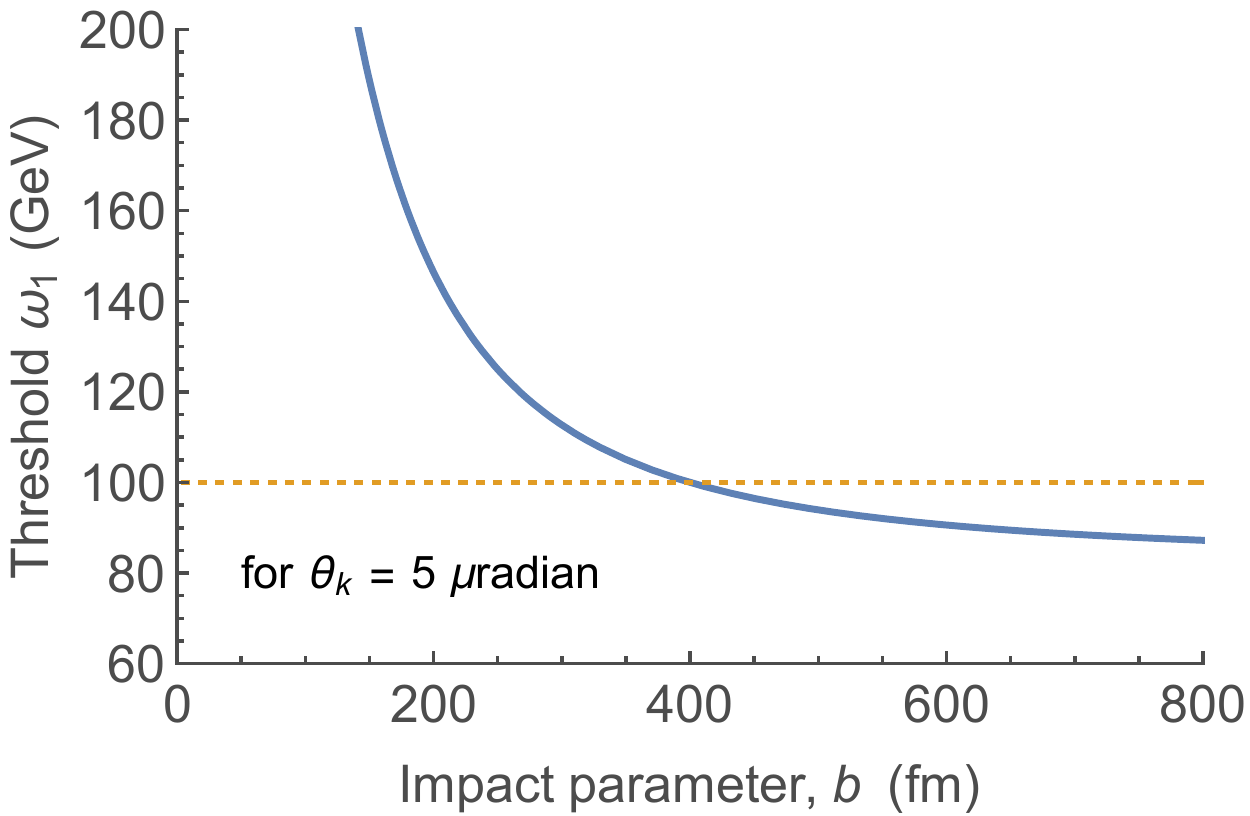}  }

\vskip 5 mm
\centerline{(b)}

\centerline{  \includegraphics[width = 0.92\columnwidth]{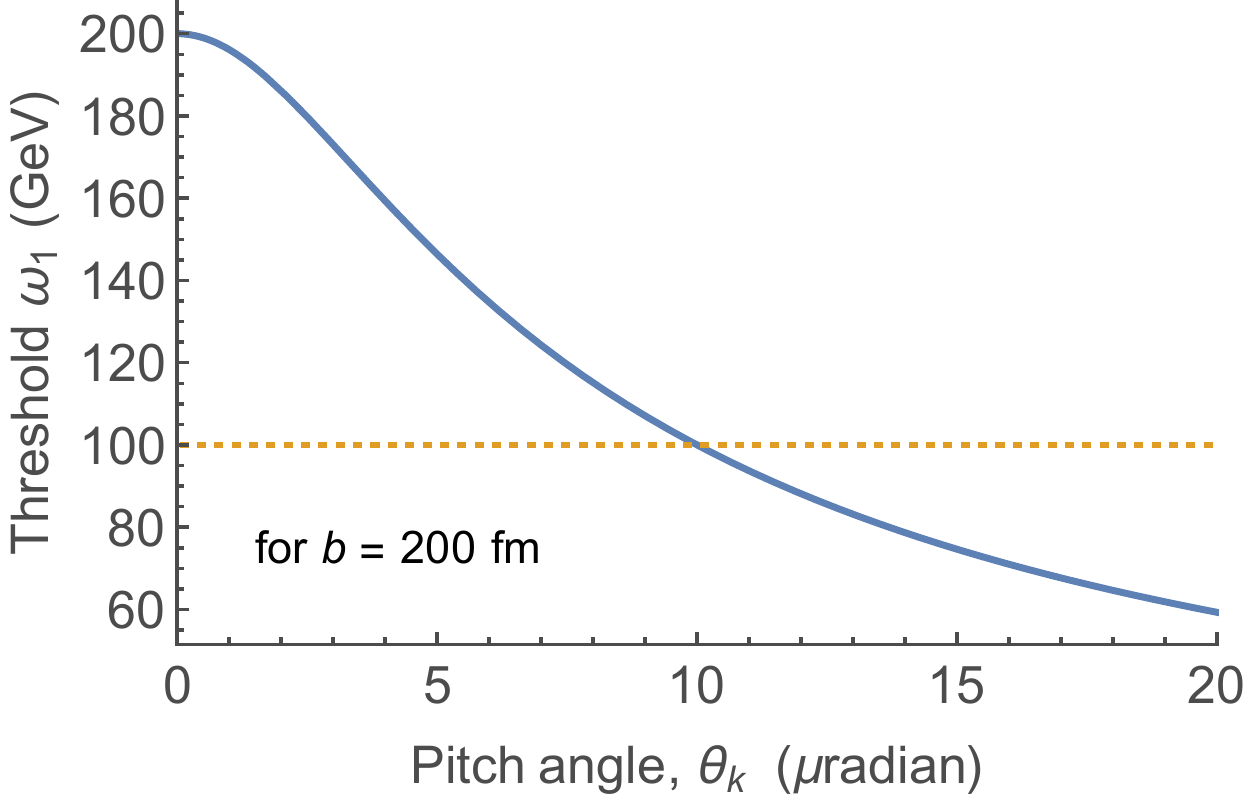}  }
\caption{Threshold energies for the energetic twisted photon in $\gamma\gamma\to e^+ e^-$ when the other photon is a visible-range background photon.  For both parts, the orange dotted line shows the threshold when both photons are plane waves.  (a) shows threshold energies as a function of impact parameter for fixed pitch angle $\theta_k = 5$ microradians, and (b) shows threshold energies as a function of pitch angle for fixed impact parameter $b = 200$ fm.}
\label{fig:astrothreshold}
\end{center}
\end{figure}

Solving the full equation for fixed $\theta_k$ shows results for threshold $\omega_1$, in contrast to the atomic or nuclear cases at nonrelativistic energies, both above and below the plane wave value: above for small impact parameter and below for large impact parameter.  Similar remarks follow for solving at varying $\theta_k$ and fixed $b$.    Plots are shown in Fig.~\ref{fig:astrothreshold} for $\ell_\gamma = 1$ and selected $\theta_k$ and $b$.  The crossover  is roughly at
\be
\label{eq:crossover}
b  \theta_k = 2 \text{ picometer}\cdot\mu\text{radian}		.
\ee
For a given pitch angle $\theta_k$, impact parameters $b$ smaller than given by the above relation will require more energy than for a plane wave to make the reaction happen, giving a decrease in the overall reaction rate.  Correspondingly, larger impact parameters require less energy and would lead to more photons interacting and being taken out of the beam.

If we considered an impact parameter at or below the wavelength of the incoming photon, we would find remarkable recoil energies.  But very small impact parameters could be much ameliorated by the finite production region of  the $e^+ e^-$ pair, so let us estimate possible parameters of the twisted VHE photon wave packet that would lead, for example, to a ten-fold increase of threshold energy of $\gamma\gamma\to e^+ e^-$ reaction, to about 1 TeV.    For small pitch angles and Eq.(\ref{eq:ggRecoil}) ten-fold energy increase would require transverse recoil of $p_T=6m_e$, which for $\ell_\gamma=1$ corresponds to an impact parameter $b=33$ fm. Remembering that the transverse profile of the twisted-photon has a bulls-eye pattern with an empty center, we can fit the position-space wave function $\psi_\gamma(\rho)$ to Bessel-Gauss profile as in Ref.~\cite{Afanasev_2018}, where the largest term is proportional to
\begin{equation}
 \psi_\gamma(\rho)=A\ J_{\ell_\gamma}({2\pi}{\rho}\sin(\theta_k) / \lambda)  \exp(-\rho^2/w_0^2),
\end{equation}
where $\rho$ describes transverse position, $J_{\ell_\gamma}$ is Bessel function of ${\ell_\gamma}$ order,  $w_0$ constrains transverse spread of the wave function at large transverse distances, and $A$ is a normalization constant.
One can get the wave function peaking at $\rho=33$ fm, to provide ten-fold increase of threshold energy, with a $\theta_k\approx 5 \times10^{-6}$ radian and $w_0\approx 60$ fm.

The detailed modification of partial-wave amplitudes and cross sections for $\gamma\gamma\to e^+e^-$ process depends significantly on the twisted-photon wave function parameters, and also on the production region of the lepton pair, and will be a subject of a separate study.

As in the case of deuteron photo-disintegration discussed previously, modification of threshold energy only applies to transitions with excess of AM coming from the twisted photon. AM of relative electron-positron motion above threshold could partially absorb AM excess of the twisted photons, but the excess recoil would persist at higher impact parameters. We also note that production of para-positronium ($p$Ps), a singlet $^1S_0$ spin state of $e^+e^-$ system, would be kinematically forbidden for a broad energy range of VHE  twisted photon.

We conclude this section with an assertion that the universe could be more transparent to twisted VHE gamma rays than to plane wave gamma rays. The result emphasizes importance of studying QED and nuclear processes with twisted gamma rays at lower energies in laboratory conditions. Such twisted gamma-sources can be built using inverse Compton scattering of high-energy electrons on the twisted optical photons \cite{Jentschura_2011,Petrillo_2016}. Possible origins of twisted gamma-rays in the astrophysics environment were discussed in the literature \cite{Taira_2018,Katoh_2017,2019NatSR...9...51M,Maruyama_2019,Maruyama:2019bin,Tamburini_2011}. 


\section{Summary}


We have demonstrated that twisted-photon absorption in several examples of quantum processes leads to additional recoil momentum of the final particles.   Overall, it leads to increased threshold energy required for the process to occur.  The increase of threshold energy depends on the impact parameter, or the location of the interaction region with respect to the vortex axis of a twisted particle,  which implies sub-wavelength position resolution for corresponding measurement. 

The excess recoil is small albeit measurable for atomic and nuclear photo-induced reactions at eV and MeV energies, and there is the possibility dramatic effects for VHE cosmic rays. 
Our findings warrant more detailed studies of twisted gamma-ray interactions in the lab, as well as new approaches to measure directly orbital AM of high-energy cosmic rays.


\section*{Acknowledgements}
The work of A.A. was supported by US Army Research Office Grant W911NF-19-1-0022.  C.E.C.~thanks the National Science Foundation (USA) for support under grant PHY-1812326. 
A.A. thanks Michael Berry for stimulating discussions of the ``superkick" during ICOAM'17 conference. Discussions with Ferdinand Schmidt-Kaler and Christian Schmiegelow on physics of cold trapped ions are also gratefully acknowledged.


\bibliography{superkick}

\begin{thebibliography}{36}
\bibitem [{cur()}]{curvenote}%
  \BibitemOpen
  \href@noop {} {}\bibinfo {note} {The probabilities $w(m_f)$ are proportional
  to the transition amplitudes that can, for example, be found in Eq.~(6) or
  Eq.~(18) of~\cite{Afanasev_2020}, and normalized so they sum to unity. (For
  Figs.~\ref{fig:Lz1} and~\ref{fig:Lz-1} when $| m_\gamma | > l_f$, the $b=0$
  point (only) should be omitted, since all transition amplitudes are zero in
  that case. For $b \ne 0$, all curves are valid, and for $| m_\gamma | \le
  l_f$ the $b=0$ point is also valid.)}\BibitemShut {NoStop}%
\end{thebibliography}

\end{document}